\documentclass[prl,fleqn,amssymb,floats,showpacs,twocolumn]{revtex4}
\usepackage{graphicx}

\newcommand{\be}{\begin{equation}}
\newcommand{\ee}{\end{equation}}
\newcommand{\bea}{\begin{eqnarray}}
\newcommand{\eea}{\end{eqnarray}}
\newcommand\egal{&\!\!\!=\!\!\!&}

\def\A{{\cal A}}
\def\D{{\cal D}}
\def\E{{\cal E}}
\def\G{{\cal G}}
\def\T{{\cal T}}
\def\b{\beta}
\def\ra{\rangle}
\def\rad{\ra\!\ra}
\def\half{{\textstyle {1 \over 2}}}
\def\sp{{\textstyle {1 \over 16}}}

\begin{document}
\draft

\title{Kramers-Wannier dualities via symmetries}
\author{Philippe Ruelle\thanks{Chercheur qualifi\'e FNRS}}
\address{Universit\'e catholique de Louvain\\
Institut de Physique Th\'eorique\\
B-1348 \hskip 0.5truecm Louvain-la-Neuve, Belgium
}
\date{\today}
\widetext
\begin{abstract}
Kramers-Wannier dualities in lattice models are intimately connected
with symmetries. We show that they can be found directly and
explicitly from the symmetry transformations of the boundary states in the
underlying conformal field theory. Intriguingly the only models with
a self-duality transformation turn out to be those with an auto-orbifold
property.
\end{abstract}

%Keywords: order-disorder dualities, symmetries, boundary conformal field theory.
\maketitle

%%%%%%%%%%%%%%%%%%%%%%%%%%%%%%%%%%%%%%%%%%%%%%%%%%%%%%%%%%%%%%%%%%%%%%%%%%%%%%

\noindent
%%%%
Universal properties of two-dimension\-al critical phenomena, when conformally
invariant, are described by conformal field theories. Virtually all their aspects
can be understood within an appropriate conformal theory. Duality transformations,
which have played a prominent role in the history of critical phenomena, have however
so far largely remained outside the conformal description. 

The notion of duality is inspired by the duality discovered by Kramers-Wannier
in the Ising model \cite{kw}, and more specifically its interpretation in terms of
disorder lines given by Kadanoff and Ceva \cite{kc}: a duality exchanges
order with disorder fields, where the disorder fields are associated with
defects lines, non-local perturbations concentrated on lines. 
The Kramers-Wannier duality had the surprising and important feature of relating the
high and low temperature regimes. Dualities were subsequently discovered in many other
models, and also generalized to relate different models \cite{sdw}.

Recently the dualities in self-dual models have been reconsidered from a conformal
theory point of view \cite{ffrs}. It was shown there that the information about the
dualities is in fact contained in the fusion algebra of special fields associated
with conformal defects.

%%%%
In this letter, we provide an alternative point of view by connecting the
dualities directly with the symmetries of the underlying conformal
theory. We give a simple and completely explicit way to find out when there is a
duality, and to compute the duality 
transformations. Although the method is general, we will mostly deal with the unitary
Virasoro minimal theories. We will conclude that only six of them have a non
trivial duality. 

The exact way the present work relates to \cite{ffrs} is not clear to us.
The dualities discussed in \cite{ffrs} are generated by specific duality
defects which are not group-like. In contrast, our approach is based on the
transformation laws of the boundary states under internal symmetries. The results 
nonetheless seem to fully agree, which indicates that the two approaches are
complementary.

%%%%
The dualities discussed here are derived in the conformal setting, namely
at criticality. They however extend at least to the perturbative neighbourhood of the
critical point, where the conformal picture holds. When the critical point is self-dual,
the duality is a true symmetry which establishes relations among correlators and
operator product coefficients. The duality is also useful to connect different
relevant perturbations. If the perturbing term is $\lambda \int \phi(x)$ for a
duality odd field, the duality establishes a correspondence between the $\lambda>0$ and
the $\lambda<0$ regimes, and between the corresponding renormalization group flows,
generalizing the duality between the low and high temperature phases of the Ising 
model.

%%%%%%%%%%%%%%%%%%%%%%%%%%%%%%%%%%%%%%%%%%%%%%%%%%%%%%%%%%%%%%%%%%%%%%%%%%%%%%

{\it The Ising model.} - It is instructive to reconsider the Ising model at 
zero magnetic field. On an $L \times T$ cylindrical lattice with free boundary 
conditions along the bottom and top boundaries, the contour representation 
of the partition functions is
\be
Z_{P,ff}^{L \times T}(\b) = 2^{LT}  \, (\cosh{\b})^{2LT-L} \, \sum_{C} \;
(\tanh{\b})^{|C|},
\ee
where the sum runs over closed contours. A contour is a set of loops drawn
on the lattice bonds, where each bond is used at most once ($|C|$ is
the number of bonds of $C$).

A contour $C$ defines the variations of a dual spin configuration $s^*$ on the 
dual lattice \cite{syo}. The dual spins are constant inside the loops of $C$, and
change sign when one crosses a contour line. If the dual spin at a single dual
site is fixed, the whole dual configuration is unambiguously fixed by $C$. 
By relating the energy of the dual configuration to the value of $|C|$, one can
write the above partition function in terms of a partition function for the
dual spins. This dual partition function is evaluated at the
dual temperature $\b^*$ defined by $\tanh{\b} = {\rm e}^{-2\b^*}$, while 
the dual spins are subjected to dual boundary conditions, usually distinct
from the original ones. In the present case, the relation reads
\bea
Z_{P,ff}^{L \times T}(\b) \egal 2^{L/2} \, (\sinh{2\b})^{LT-L/2} \times \nonumber\\
&& \hspace{5mm} \Big[ Z_{P,++}^{L \times (T+1)}(\b^*) + Z_{P,+-}^{L \times (T+1)}(\b^*) \Big],
\label{freeP}
\eea
where $Z_{P,++}$ (resp. $Z_{P,+-}$) is the cylinder partition function for fixed 
and equal (resp. opposite) spins on the two boundaries.

The same calculation can be repeated for antiperiodic boundary condition along
the perimeter of the cylinder. In this case, the energy is computed as
before except that two neighbouring spin columns are coupled
antiferromagnetically. The duality relation reads in this case
\bea
Z_{A,ff}^{L \times T}(\b) \egal 2^{L/2} \, (\sinh{2\b})^{LT-L/2} \times
\nonumber\\
&& \hspace{5mm} \Big[ Z_{P,++}^{L \times (T+1)}(\b^*) - Z_{P,+-}^{L \times 
(T+1)}(\b^*) \Big]. \label{freeA}
\eea

At the critical, self-dual point, specified by $\sinh{2\b}=1$, the 
relations (\ref{freeP}) and (\ref{freeA}) imply the following identities among 
universal partition functions,
\be
Z_{f|f} = Z_{+|+} + Z_{+|-}, \qquad Z^A_{f|f} = Z_{+|+} - Z_{+|-}.
\label{cont}
\ee
They not only show that the fixed boundary conditions are dual to the free
condition, but also emphasize the role of the symmetry group.

A toric lattice can also be considered. The contours 
break up into four subclasses, according to the number, even or odd, of non
contractible loops running in the two directions. Interestingly the duality 
relation takes the form 
\bea
Z_{PP}(\b) \egal {1 \over 2} \, (\sinh{2\b})^{LT} \times \nonumber\\
&& \hspace{-6mm} \Big[Z_{PP}(\b^*) + Z_{PA}(\b^*) + Z_{AP}(\b^*) +
Z_{AA}(\b^*)\Big].
\label{torus}
\eea
%One obtains three similar relations for $Z_{PA}, Z_{AP}, Z_{AA}$, the only
%change being signs in the linear combination. 
At the critical point, these
relations show that the Ising model is its own orbifold. (When a 
conformal theory has a symmetry group, the orbifold construction produces
another conformal theory by rearranging the fields of the various sectors \cite{DFMS}.
For the minimal models, this construction has been worked out in their restricted 
solid-on-solid lattice realization, and off criticality in \cite{fg}.)

%Let us mention that another type of fixed boundary condition, with
%alternating $+$ and $-$ spins, has been considered by Brascamp and Kunz
%\cite{bk}. A similar analysis shows that the
%dual of the BK boundary is a free boundary with an imaginary external field
%$h=i{\pi \over 2}$. Although they look completely different, the results of
%\cite{ioh} show that the BK boundary condition is in the universality class
%of the free boundary condition.

%%%%%%%%%%%%%%%%%%%%%%%%%%%%%%%%%%%%%%%%%%%%%%%%%%%%%%%%%%%%%%%%%%%%%%%%%%%%%%

{\it Duality in the conformal theory.} -
The conformal description of a system with a boundary starts with
conformally invariant boundary conditions. These can be written
as $|\alpha\ra = \sum_{i \in {\cal E}} \, c_\alpha^i \, |i\rad$, where $|i\rad$
are unphysical boundary states called Ishibashi states and the $c_\alpha^i$ are
numerical coefficients \cite{bcft}. The set ${\cal E} = {\cal E}_e \cup {\cal
E}_g \cup \ldots$ is a set of labels for all the scalar bulk fields of the
theory, {\it including those contained in sectors twisted by elements of the
symmetry group} \cite{pr,rv}. We will see that the Ishibashi states from the
twisted sectors play a central role in our derivation of the dualities.

Again the case of the Ising model is particularly instructive. There are three
physical conformal boundary conditions, the free condition $|f\ra$, and the
fixed ones $|+\ra$ and $|-\ra$. The periodic sector contains three scalar
fields, the identity, the spin field $\sigma$ and the energy density
$\varepsilon$, of chiral conformal weight 0, $1 \over 16$
and $1 \over 2$ respectively. They lead to three Ishibashi states
$|0\rad_P, \, |{1 \over 16}\rad_P$ and $|{1 \over 2}\rad_P$. The second,
antiperiodic sector contains a single scalar field, the disorder field $\mu$,
with the same conformal weight $\sp$ as the spin field, and gives rise to one
Ishibashi state $|\sp \rad_A$.

Inserting the values of the coefficients $c_\alpha^i$ \cite{bcft,pr}, we obtain the
following suggestive expansions,
\bea
|+\ra \egal {1 \over \sqrt{2}} \, \Big[ |0\rad_P + \sqrt[4]{2}\, |\sp \rad_P +
|\half \rad_P \Big], \label{+} \\
|f\ra \egal \quad \;\;\: \Big[ |0\rad_P + \sqrt[4]{2}\, |\sp \rad_A - |\half
\rad_P \Big], \label{f} \\
|-\ra \egal {1 \over \sqrt{2}} \, \Big[ |0\rad_P - \sqrt[4]{2}\, |\sp \rad_P +
|\half \rad_P \Big]. \label{-}
\eea
These equations reflect the transformation laws under the symmetry group $Z_2$,
but also reveal the duality transformations. The duality exchanges $\sigma$
and $\mu$, and therefore the states $|\sp \rad_P$ and $|\sp \rad_A$ which
are built on them. This results in the exchange of $|f\ra$ and $|+\ra$ or $|-\ra$
($|\sp \rad_A$ is defined up to a phase, a sign in particular) provided the energy
density is odd (it defines the thermal perturbation, and
%%%%
so changes sign under the low-high temperature duality). In fact, the relations
(\ref{cont}) suggest to write the duality relations as $|f\ra_P \leftrightarrow {|+\ra
+ |-\ra \over \sqrt{2}}$ and $|f\ra_A \leftrightarrow {|+\ra - |-\ra \over
\sqrt{2}}$, where $|f\ra_{P,A}$ denote the projections of $|f\ra$ on the $P$ and $A$
sectors. These two dualities are consistent with the symmetry. Together they imply
that $|f\ra = |f\ra_P + |f\ra_A$ is exchanged with $|+\ra$ (or indeed $|-\ra$).

%%%%
We now proceed to generalize these observations. 
The unitary minimal Virasoro models are classified by pairs $(\A_{2m},\G)$
with $\G$ a simply-laced Lie algebra with Coxeter number $q=2m$ or $2m+2$,
and $m \geq 1$ an integer \cite{ciz}. All the required data are encoded in the
Dynkin diagrams of $\A_{2m}$ and $\G$ \cite{bcft,pr,rv}.

The boundary states $|\alpha\ra = |(a,b)\ra = \sum_{i \in {\cal E}} \, c_\alpha^i
\, |i\rad$ can be labelled by a node $a=1,2,\ldots,m$ of the diagram $\T_m =
\A_{2m}/Z_2$ (tadpole diagram with $m$ nodes), and a node $b$ in the diagram of
$\G$. The model $(\A_{2m},\G)$ possesses an internal symmetry group $G$ equal
to the automorphism group of $\G$, and moreover the action of $G$ on the
boundary states coincides with its action on the diagram $\G$, {\it i.e.}
$^g|(a,b)\ra = |(a,g(b))\ra$. In terms of the Ishibashi states $|i\rad$, this
action is induced by the action of $G$ on the scalar fields $i \in {\cal E}$ of
all sectors. The set $\cal E$ itself decomposes as ${\cal E} = \cup
_{g \in G} \, {\cal E}_g$, where ${\cal E}_e$ corresponds to the periodic
sector. Each subset ${\cal E}_g$ labels the scalar fields contained in the sector
twisted by $g$, and equals the set of Kac labels $(r,s)$ where $r$ and $s$ run over
the exponents of $\A_{2m}$ and of $\G^g$ respectively (up to the symmetry of the
Kac table). Here $\G^g$ is the part of the diagram
of $\G$ that is left fixed by $g$, and is itself a Dynkin diagram. Finally the
coefficients $c_\alpha^i$ are explicitly known in terms of the eigendata
of the fused adjacency matrices of $\A_{2m} \times \G^g$. In particular, and
this will be crucial for what follows, if $i \in {\cal E}_g$ refers to a scalar
field in the sector twisted by $g$, then $c_\alpha^i = 0$ if $\alpha$ is not
invariant under $g$. F.i. in the Ising model $(\A_2,\A_3)$, $|f\ra = |(1,2)\ra$
(first node of $\T_1$, second of $\A_3$) is the only $Z_2$ invariant boundary state,
and thus the only one to have a projection on $|\sp\rad_A$.

We will impose three basic requirements on the dualities: (i) a duality must
exchange bulk fields from different sectors, since otherwise the transformation would
be called a symmetry rather than a duality; (ii) it must be invertible; and
(iii) it has to be consistent with both the chiral algebra and the internal symmetry
(necessarily non-trivial). 

%%%%%%%%%%%%%%%%%%%%%%%%%%%%%%%%%%%%%%%%%%%%%%%%%%%%%%%%%%%%%%%%%%%%%%%%%%%%%%

{\it Unitary minimal conformal models and relatives.} -
The diagonal theories $(\A_{2m},\A_{q-1})$ all have a $Z_2$ symmetry. The
boundary conditions are parametri\-zed by $a$ in $\{1,2,\ldots,m\}$ and $b$ in
$\{1,2,\ldots,q-1\}$, among which those with $b={q \over 2}$ are
$Z_2$ invariant. We choose ${\cal E}_e = \{(r,s) \in [1,2m] \times [1,q-1]
\;:\; r \;{\rm odd}\}$ to label the scalar fields of the periodic
sector. For $i=(r,s)$ in ${\cal E}_e$, the coefficients are 
$c_{(a,b)}^{i} \sim \sin{\pi qar \over 2m+1} \, \sin{\pi
(2m+1)bs \over q}$ up to a non-zero factor.

There are two sectors, periodic and antiperiodic. Requirement (i)
means that some $|i\rad_P$ are exchanged with some $|j\rad_A$, and therefore 
implies that a boundary state $|\alpha\ra$ which is not invariant under
the $Z_2$ symmetry and which has a non-zero projection on all the states 
$|i\rad_P$, is necessarily exchanged with a boundary state
$|\alpha^*\ra$ which expands on Ishibashi states from the $A$ sector, that is, with
a $Z_2$ invariant boundary state. Consistency with the symmetry $Z_2$ actually
requires that $|\alpha^*\ra$ be dual to the combination $(|\alpha\ra +   
\:^g|\alpha\ra)/\sqrt{2}$.

Since the duality must be invertible, the number of such pairs $|\alpha\ra,
\:^g|\alpha\ra$ must be smaller or equal to the number of invariant boundary
states, equal to $m$. From the formula given above, the coefficients $c_{(a,b)}^i$
are different from zero for all $i$ in the periodic sector, if and only if $a$
is coprime with $2m+1$ and $b$ is coprime with $q$. The total number of pairs
$|\alpha\ra,\:^g|\alpha\ra$ is thus equal to ${1 \over 4} \phi(q)
\phi(2m+1)$, with $\phi(n)$ the Euler totient function, so that the
aforementioned inequality reads $\phi(q) \phi(2m+1) \leq 4m$.
The crude lower bound $\phi(n) \geq n^{3/5}$ if $n$ is odd, and $\phi(n)
\geq ({n \over 2})^{3/5}$ if $n$ is even, is enough to show that the inequality
is violated for all $m > 128$. Checking the finite number of remaining cases
leaves only four cases: $(m,q) = (1,4),\, (2,4), \,(2,6)$ and (3,6).

By looking at the expansions of the boundary conditions in terms of
the Ishibashi states, the two cases $(m,q)=(2,6),(3,6)$ are easily ruled out for
not having a consistent duality. The last two cases $(m,q)=(1,4)$ and (2,4) 
correspond to the Ising model, discussed above, and the tricritical Ising model, 
both self-dual at the critical point. 

The tricritical Ising model $(\A_4,\A_3)$ has six conformal
boundary conditions. Four of them expand as
\bea
|(1,1)\ra \egal C \Big[ |0\rad_P + \eta |\textstyle {1
\over 10}\rad_P + \eta |{3 \over 5}\rad_P + |{3 \over 2}\rad_P \nonumber\\
&& \hspace{15mm} +\; \sqrt[4]{2}\textstyle 
|{7 \over 16}\rad_P + \sqrt[4]{2} \eta |{3 \over 80}\rad_P \Big], \\
|(1,2)\ra \egal \sqrt{2} C \Big[ |0\rad_P - \eta |\textstyle {1
\over 10}\rad_P + \eta |{3 \over 5}\rad_P - |{3 \over 2}\rad_P \nonumber\\
&& \hspace{15mm} +\; \sqrt[4]{2} \textstyle 
|{7 \over 16}\rad_A + \sqrt[4]{2} \eta |{3 \over 80}\rad_A \Big], \\
|(2,1)\ra \egal C \Big[ \eta^2 |0\rad_P - \eta^{-1} |\textstyle {1
\over 10}\rad_P - \eta^{-1} |{3 \over 5}\rad_P + \eta^2 |{3 \over 2}\rad_P
\nonumber\\
&& \hspace{15mm} -\; \sqrt[4]{2} \textstyle \eta^2 |{7 \over 16}\rad_P +
\sqrt[4]{2} \eta^{-1} |{3 \over 80}\rad_P \Big], \\
|(2,2)\ra \egal \sqrt{2} C \Big[ \eta^2 |0\rad_P + \eta^{-1} |\textstyle {1
\over 10}\rad_P - \eta^{-1} |{3 \over 5}\rad_P - \eta^2 |{3 \over 2}\rad_P
\nonumber\\
&& \hspace{15mm} -\; \sqrt[4]{2} \eta^2 \textstyle |{7 \over 16}\rad_A +
\sqrt[4]{2}
\eta^{-1} |{3 \over 80}\rad_A \Big], 
\eea
where $C=\sqrt{\sqrt{5}-1 \over 8}$ and $\eta = \sqrt{\sqrt{5}+1 \over 2}$.
The last two, $|(1,3)\ra$ and $|(2,3)\ra$, are the $Z_2$ transforms of
$|(1,1)\ra$ and $|(2,1)\ra$ respectively, and are simply obtained from them
by changing the sign of the coefficients of $|{7 \over 16}\rad_P$ and $|{3
\over 80}\rad_P$, since the corresponding bulk fields are odd under the $Z_2$.

The duality transformations \cite{tim} are easily read off from these
expansions. The two bulk fields $\varepsilon = ({1 \over 10},{1 \over 10})$
and $\varepsilon'' = ({3 \over 2},{3 \over 2})$ are odd, whereas $\varepsilon'
= ({3 \over 5},{3 \over 5})$ is even; $\sigma = ({3 \over 80},{3 \over
80})_P$ is exchanged with $\mu = ({3 \over 80},{3 \over 80})_A$, and 
$\sigma' = ({7 \over 16},{7 \over 16})_P$ with $\mu' = ({7 \over 16},{7 \over
16})_A$; finally the duality between the boundary conditions reads
\bea
|(1,2)\ra_{P,A}\, &\leftrightarrow & {\textstyle {1 \over \sqrt{2}}} [|(1,1)\ra \pm
|(1,3)\ra], \\
|(2,2)\ra_{P,A}\, &\leftrightarrow & {\textstyle {1 \over \sqrt{2}}} [|(2,1)\ra \pm
|(2,3)\ra].
\eea

The complementary models $(\A_{2m},\D_{{q \over 2}+1})$ have a $Z_2$ internal
symmetry for $q \geq 8$, and an $S_3$ symmetry if $q=6$.

Assume first $q \geq 8$. There are again two sectors, P and A, and
correspondingly two sets of exponents, 
\bea
{\cal E}_e \egal \{(r,s) \in [1,2m] \times [1,q-1] \;:\; r,s \;{\rm odd}\} 
\nonumber\\
&& \hspace{2cm} \cup\; \{(r,\textstyle {q \over 2}) \;:\; r \; {\rm odd} \in
[1,2m]\},
\label{ee}\\
{\cal E}_g \egal \{(r,s) \in [1,2m] \times [1,q-1] \;:\; r \; {\rm odd}, \,
s \: {\rm even}\}. \label{eg}
\eea

As before an essential requirement is that the duality exchanges Ishibashi
states from the two sectors, $|i\rad_P \leftrightarrow |j\rad_A$. As it must
also preserve the scaling dimensions, this actually requires $i=j$ and so 
${\cal E}_e \cap {\cal E}_g$ must be non-empty. That alone excludes the cases
$q=2 \bmod 4$. 

In addition, the inclusion ${\cal E}_g \subset {\cal E}_e$ should hold for a
duality to exist. Assume the
contrary. The states $|j \rad_A$ for those $j$ which are not in ${\cal E}_e$
should be left fixed by the duality, and so the $Z_2$ invariant physical
boundary states which have a non-zero projection on them would be permuted
among themselves. The values of the coefficients show that this is the
case of all invariant states. The duality would then mix the $Z_2$ invariant
states among themselves, and being invertible, would also mix the non-invariant
states among themselves, contradicting the exchange of some $|i\rad_P$
with some $|j\rad_A$. 

A simple look at the sets (\ref{ee}) and (\ref{eg}) shows that no value of $q
\geq 8$ satisfies the condition ${\cal E}_g \subset {\cal E}_e$.
The remaining two models $q=6$, namely the critical and tricritical 3-Potts
models, are the only ones to qualify for a non-trivial duality, and easily 
checked to have one. The nature of these dualities is somewhat different
from that in the Ising model, as they are induced by the $Z_3$ subgroup of a
larger symmetry group $S_3$. 

In order to treat the two models simultaneously, and also to lighten the
expressions, we will compute their duality in terms of their parent 
theory through the coset construction, namely the $\D_4$ model with affine symmetry
$\widehat{su}(2)$ at level $k=4$. The way this model descends onto the
two 3-Potts models $(\A_4,\D_4)$ (critical) and $(\A_6,\D_4)$ (tricritical) is
standard and yields the results in a straightforward fashion.

The $\D_4$ model has an $S_3$ internal symmetry, for which
all twisted torus and cylinder partition functions have been computed in
\cite{rv}. It has four physical boundary conditions, labelled by the nodes
of the $\D_4$ Dynkin diagram, with the state $|4\ra$ being fully invariant under
$S_3$. The other three are cyclically rotated by the $Z_3$ subgroup, and
each one is left invariant by a $Z_2$ subgroup. We call $r$ the generator of the
$Z_3$ subgroup, and $g,g',g''$ the generators of the three conjugate $Z_2$
subgroups.

The full expansions in terms of the affine Ishibashi states read \cite{rv}
\bea
|1\ra \egal {1 \over \sqrt[4]{3}} \Big\{ |1\rad_e + |3\rad_e + |3'\rad_e +
|5\rad_e \nonumber\\
&& \hspace{30mm} + \; \sqrt[4]{3} (|2\rad_g + |4\rad_g)\Big\},\\
|2\ra \egal {1 \over \sqrt[4]{3}} \Big\{ |1\rad_e + \omega \, |3\rad_e +
\omega^2 \, |3'\rad_e + |5\rad_e \nonumber\\
&& \hspace{30mm} + \; \sqrt[4]{3} (|2\rad_{g'} + |4\rad_{g'})\Big\},\\
|3\ra \egal {1 \over \sqrt[4]{3}} \Big\{ |1\rad_e + \omega^2 \, |3\rad_e +
\omega \, |3'\rad_e + |5\rad_e \nonumber\\
&& \hspace{30mm} + \; \sqrt[4]{3} (|2\rad_{g''} + |4\rad_{g''})\Big\},\\
|4\ra \egal \sqrt[4]{3} \, \Big\{ |1\rad_e + |3\rad_r + |3\rad_{r^2} -
|5\rad_e \nonumber\\
&& \hspace{-5mm} +\; {1 \over \sqrt[4]{3}} (|2\rad_g + |2\rad_{g'} +
|2\rad_{g''} - |4\rad_g - |4\rad_{g'} - |4\rad_{g''})\Big\}, \nonumber \\
\eea
where $\omega \neq 1$ is a third root of unity. 

One deduces the following duality transformations: the periodic fields $3_e$
and $3'_e$ are exchanged with the twist fields $3_r$ and $3_{r^2}$, whereas
$5_e$ is odd; in the three pairs $(2_g,4_g)$, $(2_{g'},4_{g'})$ and
$(2_{g''},4_{g''})$, the fields with label 2 are even and those with label 4
are odd; for the boundary conditions, one has
\bea
|4\ra_e &\leftrightarrow& (|1\ra_e + |2\ra_e + |3\ra_e)/\sqrt{3}, \quad
|4\ra_g \leftrightarrow |1\ra_g, \\
|4\ra_r &\leftrightarrow& (|1\ra_e + \omega|2\ra_e + \omega^2 |3\ra_e)/\sqrt{3}, 
\eea
up to symmetries. 
%These transformations imply by the identities among cylinder
%partition functions $Z_{4|4} = Z_{1|1} + Z_{1|2} + Z_{1|3}$, $Z_{4|4}^{(r)} =
%Z_{1|1}^{} + \omega Z_{1|2}^{} + \omega^2 Z_{1|3}^{}$, and $Z_{4|4}^{(g)} =
%Z_{1|1}^{(g)}$, where $Z^{(g)}$ and $Z^{(r)}$ are cylinder partition functions with a
%$g$ or $r$ twist line running from one boundary to the other one. 

The critical Potts model is simply obtained by juxtaposing an $r=1,2$ index to
the above $su(2)$ labels to form a Kac label. The known duality transformations
are then recovered \cite{3potts}. The duality of the tricritical 3-Potts model is
similar. 

The last unitary Virasoro minimal models with a symmetry ($Z_2$) are
the two models $(\A_{10},\E_6)$ and $(\A_{12},\E_6)$. They both possess a non
trivial duality which comes directly from the duality of the
$\E_6$ model with affine symmetry $\widehat{su}(2)$ at level $k=10$. 

To conclude, only six unitary minimal models have a duality, and they are
precisely the only six models to be their own orbifold. They all 
inherit their duality from that of the parent theories, the $\A_3, \D_4$ and
$\E_6$ models with an $\widehat{su}(2)$ affine symmetry, which again are the
only three $\widehat{su}(2)$ affine theories to be self-orbifold. 
%A similar analysis can be carried out for theories with an $\widehat{su}(3)$ affine
%symmetry
%, for which the internal symmetries, the twisted torus and cylinder
%partition functions have been computed in 
%\cite{rv}. A single model has an
%auto-orbifold property, namely the self-conjugate $\E_{12}$ model, under
%its $Z_3$ symmetry. 
%The corresponding theory with boundaries has not been
%clarified yet as there are (at least) three isospectral graphs that may be used
%to define it \cite{bcft}. 
%We have checked 
%however 
%that only the third graph $\E_{12}^{(3)}$ \cite{bcft} defines a theory with a 
%non-trivial duality.

The reference \cite{fg}, which generalizes the torus identity (\ref{torus}) to
all unitary minimal models, strongly points to a general relation between a
duality and the orbifold construction. The above observations
can be repeated for pairs of models which are orbifold of each other, and lead
to duality relations. For instance, the orbifold of the $\D_4$ model with
affine symmetry $\widehat{su}(2)_4$ under one of its $Z_2$ subgroup is the
$\A_5$ model (same level). One finds the duality relations 
\bea
&& {1 \over \sqrt{2}} (|1\ra_\A + |5\ra_\A) \leftrightarrow |1\ra_\D\,, \quad
{1 \over \sqrt{2}} (|2\ra_\A + |4\ra_\A) \leftrightarrow |4\ra_\D\,, \nonumber\\
&& |3\ra_\A \leftrightarrow {1 \over \sqrt{2}} (|2\ra_\D + |3\ra_\D)\,,
\eea
between the boundary conditions of the two models.

%%%%%%%%%%%%%%%%%%%%%%%%%%%%%%%%%%%%%%%%%%%%%%%%%%%%%%%%%%%%%%%%%%%%%%%%%%%%%%

\acknowledgments

I would like to thank the authors of \cite{ffrs} for useful discussions, and for 
communicating as yet unpublished results, and Paul Fendley for drawing my 
attention to the reference \cite{fg}. I thank the Belgian Fonds National 
de la Recherche Scientifique for financial support.

%%%%%%%%%%%%%%%%%%%%%%%%%%%%%%%%%%%%%%%%%%%%%%%%%%%%%%%%%%%%%%%%%%%%%%%%%%%%%%

\end{document}